\documentclass[conference]{IEEEtran}
\usepackage{amsmath, bm}
\usepackage{gensymb}
\usepackage{xcolor}
\usepackage{colortbl}
\usepackage{lipsum}
\usepackage{multirow}
\usepackage{balance}
\usepackage{units}

\usepackage[pdftex]{graphicx}
\DeclareGraphicsExtensions{.pdf,.jpeg,.png}
\usepackage{float}
\usepackage{subcaption}
\usepackage[outdir=./Figures/]{epstopdf}
\makeatletter
\def\ps@IEEEtitlepagestyle{
  \def\@oddfoot{\mycopyrightnotice}
  \def\@evenfoot{}
}
\def\mycopyrightnotice{
  {\footnotesize
  \begin{minipage}{\textwidth}
  \centering
  \copyright~2020~IEEE. Personal use of this material is permitted. Permission from IEEE must be obtained for all other uses, in any current or future media, including reprinting/republishing this material for advertising or promotional purposes, creating new collective works, for resale or redistribution to servers or lists, or reuse of any copyrighted component of this work in other works. This paper has been accepted for publication at the 2020 IEEE International Symposium on Circuits and Systems (ISCAS).
  \end{minipage}
  }
}

\begin{document}

\title{Accurate Emulation of Memristive Crossbar Arrays for In-Memory Computing}

\author{\IEEEauthorblockN{Anastasios Petropoulos\IEEEauthorrefmark{1},
Irem Boybat\IEEEauthorrefmark{2}\IEEEauthorrefmark{3},
Manuel Le Gallo\IEEEauthorrefmark{2},
Evangelos Eleftheriou\IEEEauthorrefmark{2},\\
Abu Sebastian\IEEEauthorrefmark{2}, and
Theodore Antonakopoulos\IEEEauthorrefmark{1}}
\IEEEauthorblockA{\IEEEauthorrefmark{1}University of Patras, Dept. of ECE, 26504 Patras, Greece, Email: \{a.petropoulos, antonako\}@ece.upatras.gr}
\IEEEauthorblockA{\IEEEauthorrefmark{2}IBM Research - Zurich, 8803 R\"{u}schlikon, Switzerland, Email: \{ibo, anu, ele, ase\}@zurich.ibm.com}
\IEEEauthorblockA{\IEEEauthorrefmark{3}Ecole Polytechnique Federale de Lausanne (EPFL), 1015 Lausanne, Switzerland}
}

\maketitle

\begin{abstract}
In-memory computing is an emerging non-von Neumann computing paradigm where certain computational tasks are performed in memory by exploiting the physical attributes of the memory devices. Memristive devices such as phase-change memory (PCM), where information is stored in terms of their conductance levels, are especially well suited for in-memory computing. In particular, memristive devices, when organized in a crossbar configuration can be used to perform matrix-vector multiply operations by exploiting Kirchhoff's circuit laws. To explore the feasibility of such in-memory computing cores in applications such as deep learning as well as for system-level architectural exploration, it is highly desirable to develop an accurate hardware emulator that captures the key physical attributes of the memristive devices. Here, we present one such emulator for PCM and experimentally validate it using measurements from a PCM prototype chip. Moreover, we present an application of the emulator for neural network inference where our emulator can capture the conductance evolution of approximately 400,000 PCM devices remarkably well.
\end{abstract}
\IEEEpeerreviewmaketitle

\vspace{-0.10cm}
\section{Introduction}\label{sec:intro}
\vspace{-0.07cm}
The explosive growth in data-centric artificial intelligence related applications has necessitated the exploration of non-von Neumann computing paradigms such as in-memory computing. In in-memory computing, the physical attributes of memory devices are exploited to perform computational tasks in place without the need to shuttle around data between the memory and the processing units \cite{Y2017sebastianNatComm,Y2018ielminiNatureElectronics,Y2019vermaSSCM,serb2016unsupervised,yu2018neuro,vourkas2016emerging}. A new class of emerging memory devices known as resistive memory or memristive devices are particularly well suited for in-memory computing \cite{burr2017neuromorphic}. For example, the memristive devices, when organized in a crossbar configuration can be used to perform matrix-vector multiply operations. Here, the matrix elements are stored in terms of the conductance values of the memristive devices. By exploiting Ohm's law and Kirchhoff's current summation law, the matrix-vector multiply operation can be performed in constant time. This computational capability makes in-memory computing especially interesting for applications such as deep learning training and inference, where cascaded stages of matrix-vector multiplications form the bulk of computation \cite{Y2015burrTED,Y2018wangNatureElectronics,Y2019sebastianVLSI}. The forward propagation (inference) stage, as well as the backpropagation, can be realized by merely reading the array.
\begin{figure}[h!]
\centering
\includegraphics[width=0.95\columnwidth]{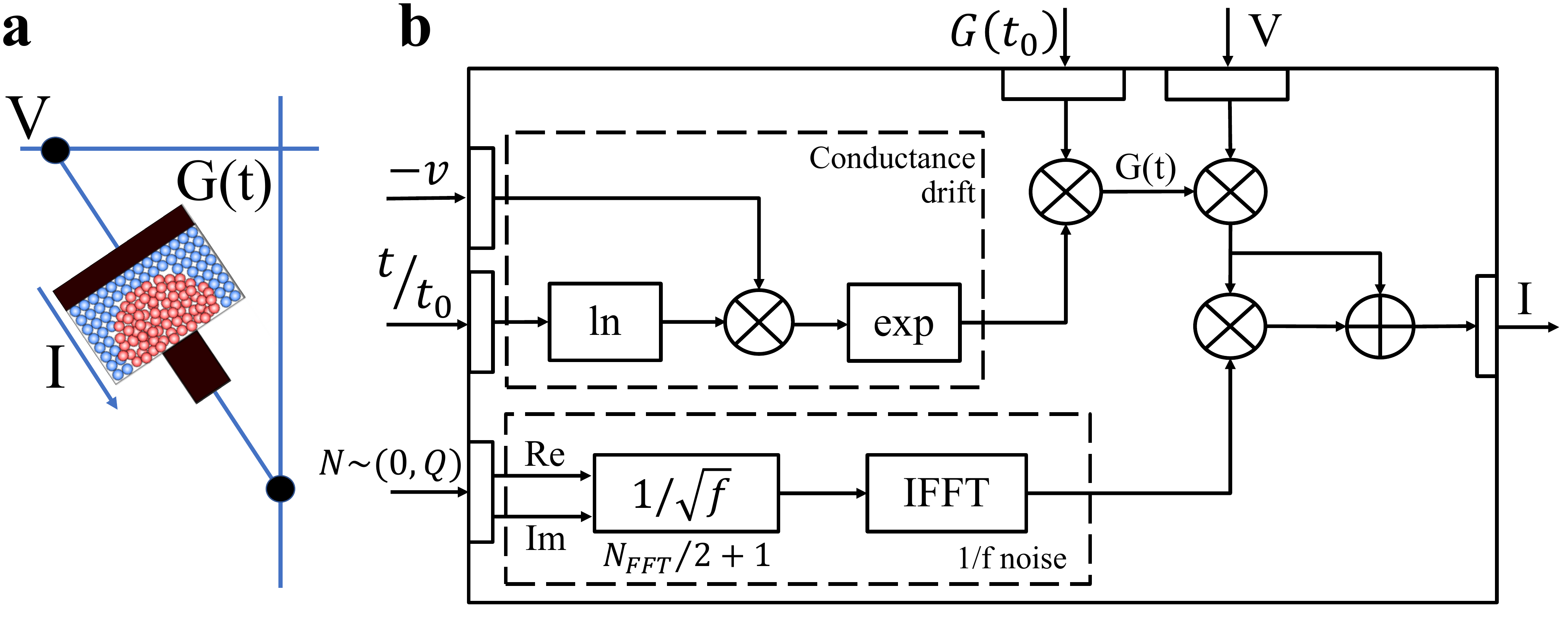}
\vspace{-0.05cm}
\caption{(a) Schematic illustration of a PCM device with a mushroom-type device geometry. (b) Corresponding PCM cell model used for the FPGA-based emulator design.} \label{fig:PCM_Cell_Model}
\vspace{-0.55cm}
\end{figure}

In spite of the promise of in-memory computing for applications such as deep learning, several open questions need to be addressed. First, it is essential to understand the computational reliability and accuracy of memristive in-memory cores for a range of applications since memristive devices exhibit non-idealities, such as temporal variations of conductance values. Identifying desired device characteristics for target applications can provide useful insight into future device designs. Furthermore, it is critical to develop efficient system architectures that involve cascaded memristive in-memory cores for applications such as deep learning. Note that compared to all-digital implementations, in-memory computing is more amenable for highly pipelined dataflows. Finally, it is of significant importance to develop a versatile software stack that can map the applications to the multi-core in-memory computing hardware. An accurate and fast hardware emulator of memristive devices and computing cores will be an indispensable tool to address all of these goals. In comparison to a software simulator, a custom-designed hardware counterpart can perform the prototyping of an in-memory computing core in a more rapid manner.
\begin{figure*}[!ht]
\centering
\includegraphics[width=0.93\textwidth]{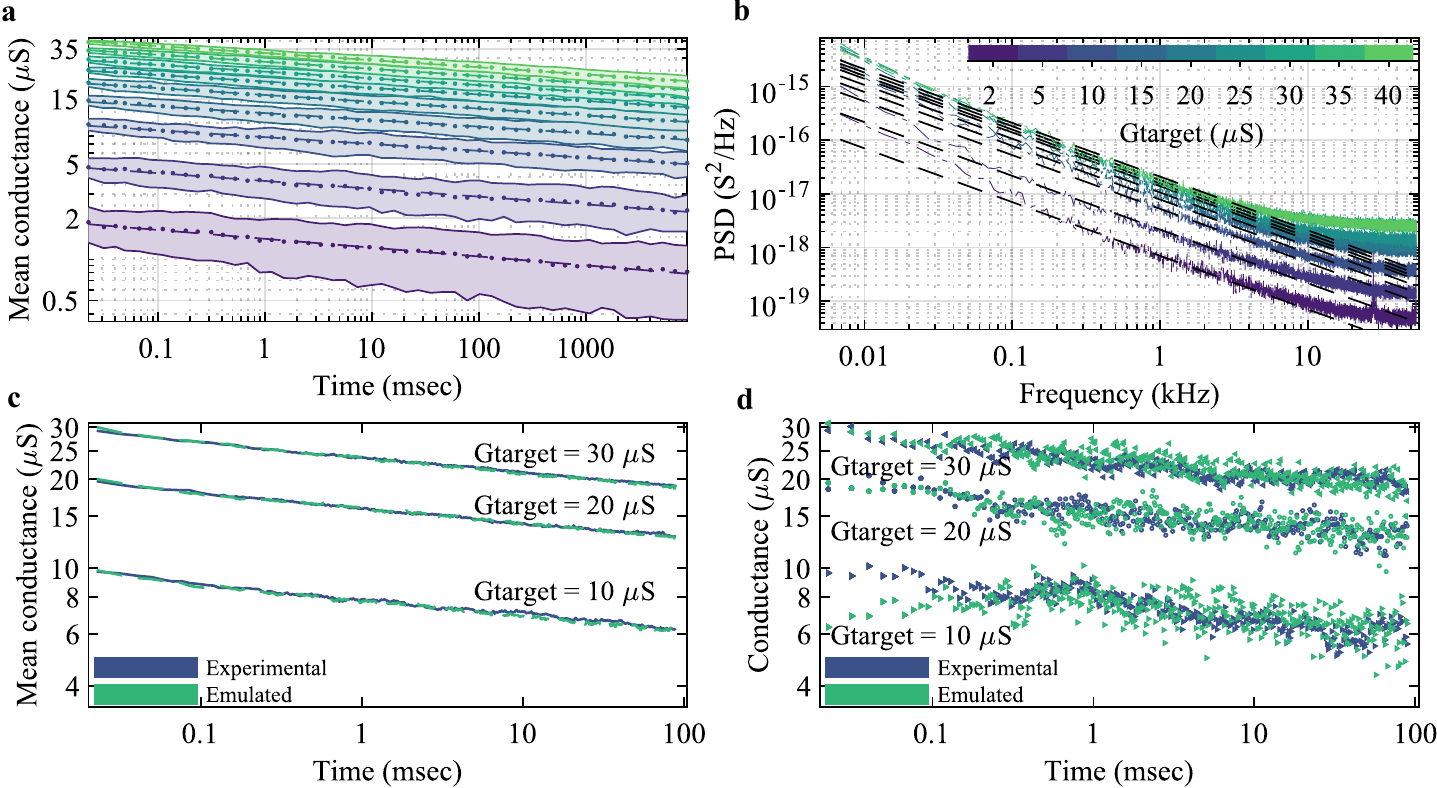}
\vspace{-0.10cm}
\caption{Experimental measurements on mushroom-type PCM devices fabricated in \unit[90]{nm} technology node. (a) Mean conductance evolution of 100 devices programmed to different target conductances and the corresponding linear fits according to Eq. (\ref{drift_module_eq}). The shades denote one standard deviation. (b) Power spectral density (PSD) of the conductance signals for each target conductance level and the corresponding fit according to Eq. (\ref{noise_eq}). (c) Emulation of the mean conductance evolution of 100 devices programmed to three different target conductance levels. (d) Emulation of the conductance evolution of 3 PCM devices.}
\label{fig:PCM_Cell_Model_Exper_Results}
\vspace{-0.55cm}
\end{figure*}

An FPGA-based hardware emulator for PCM arrays, which can mimic the temporal conductance evolution of PCM, has been previously demonstrated in \cite{petropoulos2018accurate}. The system is shown to perform a matrix-vector multiplication on a 256x256 emulated array in only 136.16 microseconds. However, functional verification with experimental data has not been demonstrated yet. In this paper, we show for the first time an FPGA-based hardware emulator that can reliably capture experimental PCM characteristics. In Section II, we present the emulation of single PCM devices in an FPGA where we capture the key physical attributes such as conductance drift and $1/f$ noise. We validate the emulator using experimental measurements from 100 devices from a prototype PCM chip programmed to various conductance levels. In Section III, we present how a PCM multi-cell crossbar array emulation can be constructed. Finally, in section IV, we illustrate the application of the PCM crossbar emulator for neural network inference, and we validate our results with an experiment involving approximately 400,000 PCM devices. 
 
\vspace{-0.10cm}
\section{PCM Cell Emulation}\label{sec:cellemul}
\vspace{-0.08cm}
PCM is arguably the most advanced resistive memory technology and has been widely employed for in-memory computing \cite{Y2016burrJETCAS, Y2019sebastianJPD, le2018mixed, boybat2018neuromorphic, nandakumar2018mixed}. PCM exploits the behavior of certain phase-change materials such as Ge$_2$Sb$_2$Te$_5$ that can be switched reversibly between amorphous and crystalline phases of different electrical resistivity. A PCM device consists of a certain volume of this phase-change material sandwiched between two electrodes (see Fig. \ref{fig:PCM_Cell_Model}a). By applying suitable electrical pulses, referred to as programming pulses, it is possible to alter the phase configuration within the PCM device and achieve different conductance values. By iterative programming schemes comprising multiple program-and-verify steps, it is possible to obtain any desired conductance value within a certain error margin \cite{papandreou2011programming}. However, the programmed conductance values exhibit temporal variations such as drift, which is attributed to the structural relaxation of the unstable amorphous phase \cite{Y2018legalloAEM}, and $1/f$ noise. These temporal variations are shown to be detrimental for PCM-based implementations \cite{joshi2019accurate} and hence need to be well captured by a PCM cell emulator.

\begin{figure*}[!t]
\centering
\includegraphics[width=1.0\textwidth,height=5.1cm,keepaspectratio]{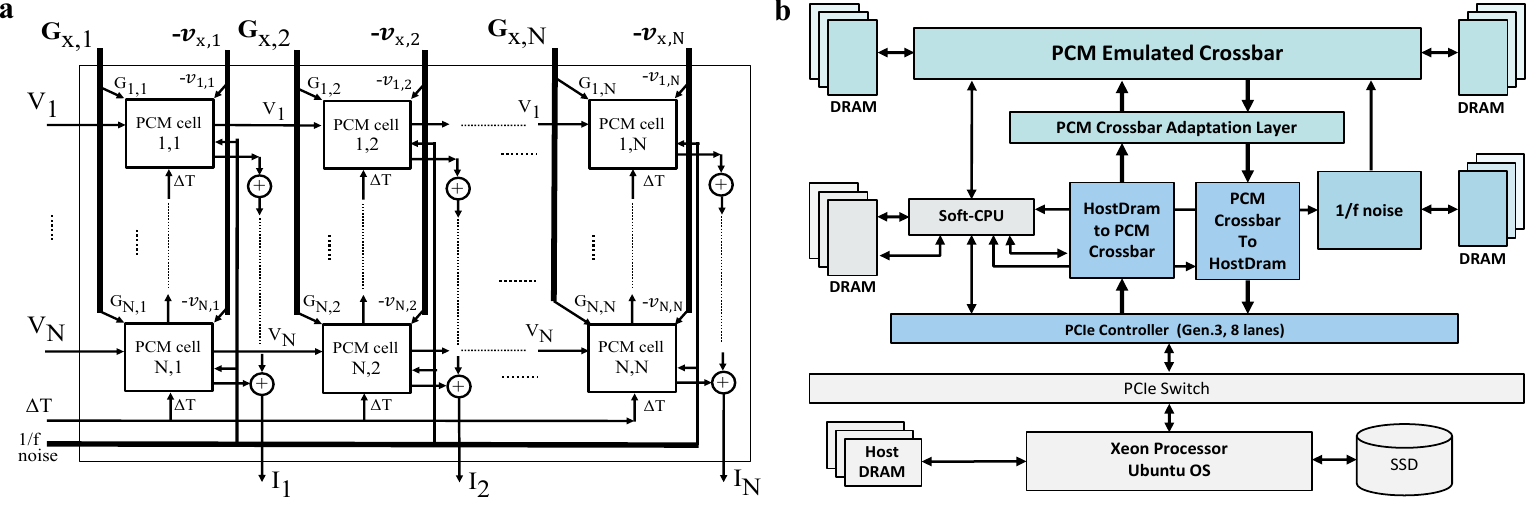}
\vspace{-0.10cm}
\caption{The PCM crossbar emulator. (a) Functional diagram \cite{petropoulos2018accurate}. (b) Hardware diagram of the PCIe-based FPGA system.} \label{fig:PCM_Crossbar}
\vspace{-0.55cm}
\end{figure*}

As shown in Fig. \ref{fig:PCM_Cell_Model}b, the PCM cell emulator consists of two functional modules, one for the conductance drift and one for the $1/f$ noise. The drift is modeled according to the following power-law equation and can be rearranged for ease of hardware implementation.
\begin{equation} \label{drift_module_eq}
G(t) = G(t_0)\left(\dfrac{t}{t_0}\right)^{-\nu} = G(t_0) exp\left(-\nu \ln\left(\dfrac{t}{t_0}\right)\right)
\end{equation}
In Eq. (\ref{drift_module_eq}), $G(t)$ denotes the conductance value at time instance $t$, $G(t_0)$ denotes the conductance at time $t_0$, and $\nu$ is the drift exponent. For \unit[90]{nm} doped GST devices, $\nu$ is reported to take values between 0.03 and 0.1, depending on the initial amorphous volume created with the programming pulse \cite{le2017compressed, boybat2018impact}. There is also variability associated with $\nu$ \cite{le2017compressed,boniardi2010statistics,nandakumar2019phase}. To capture these observations, we sample the drift exponent of individual devices from a Gaussian distribution with a certain mean and standard deviation. For the $1/f$ noise module, a hardware block was designed in order to implement the equation:
\begin{equation} \label{noise_eq}
S_{I_\text{noise}}(f) = I_\text{read}^{2}Q \dfrac{1}{f}
\end{equation}
$S_{I_\text{noise}}(f)$ denotes the power spectral density associated with the read noise \cite{close2010device}. $I_\text{read}$ denotes the mean read current when biased by the read voltage, $V$. As shown in Fig. \ref{fig:PCM_Cell_Model}b, we generate two independent and normally distributed random vectors with a dimension of $N_{FFT}/2 + 1$, with known variance, $Q$, and zero mean, and we use them as a complex Gaussian random vector in the frequency domain. The amplitude of the complex Gaussian random vector is scaled by the $1/\sqrt f$ factor. Then the negative frequency spectral samples are determined to satisfy for Hermitian symmetry. Finally, the inverse Fourier transform ($N_{FFT}$ points) is applied for generating a real-valued time series with the desired noise characteristics \cite{timmer1995generating}.

The underlying functions associated with the drift and noise functional modules were implemented with the utilization of floating-point cores that integrate DSP slices, which are pipelined for achieving high throughput, if a large number of cells has to be emulated. With this model, we can investigate the influence of drift and $1/f$ noise on scalar multiplication.

For the experimental validation of the PCM cell emulator, we used an experimental platform with 1 million mushroom-type PCM cells, with doped Ge$_2$Sb$_2$Te$_5$ as the phase-change material, and fabricated in \unit[90]{nm} CMOS technology. In order to verify our PCM cell emulator for different conductance levels, we programmed 100 devices to a range of 2 $\mu$S to 40 $\mu$S using iterative programming. Subsequently, the read current from each device measured for approximately 9 seconds with a sampling rate of 112 kHz. The device conductances were estimated based on the read voltage of $V = \unit[0.2]{V}$.

Fig. \ref{fig:PCM_Cell_Model_Exper_Results}a shows the evolution of PCM conductance states. It can be seen that the mean behavior matches the relationship predicted by Eq. (\ref{drift_module_eq}). For each targeted conductance level, a line is fitted on the average conductance evolution. The drift exponent is calculated from the slope of this linear fit. Note that $\nu$ depends on the initially created amorphous volume. We assume a constant standard deviation of 0.02 for $\nu$. To estimate the noise and its power spectral density, we used the read measurements obtained during the last second. The reason for this is to decouple the effect of drift from the $1/f$ noise measurement as drift slows down significantly with time. Fig. \ref{fig:PCM_Cell_Model_Exper_Results}b presents the PSD of the $1/f$ noise and the corresponding fitting curves with respect to Eq. (\ref{noise_eq}) for different target conductance levels. Based on these measurements, it is observed that $Q$ becomes higher as the target conductance level becomes lower, also reported by \cite{nandakumar2019phase, fantini2006experimental}. The observed values of Q were from 5.1$\times10^{-5}$ to 1.1$\times10^{-3}$. Typically, the noise in PCM follows a $1/f^{\gamma}$ relationship, where $\gamma$ is reported to be within the range of 0.9-1.1 \cite{fantini2008characterization}. The deviation from the $1/f$ behavior is also evident in Fig. \ref{fig:PCM_Cell_Model_Exper_Results}b. However, for modeling simplicity, we assume an ideal $1/f$ relationship, where we use $N_{FFT} = $ 1024 points for the inverse Fourier transform in the emulator. The extracted experimental parameters were used in the PCM cell emulator, and the device behavior was emulated. As shown in Fig. \ref{fig:PCM_Cell_Model_Exper_Results}c and Fig. \ref{fig:PCM_Cell_Model_Exper_Results}d, the emulated conductance evolution over time matches the experimental behavior remarkably well. Both the mean conductance behavior and individual device evolution are faithfully captured. 

\vspace{-0.10cm}
\section{PCM Crossbar Emulation}\label{sec:cbemul}
\vspace{-0.08cm}
\begin{figure*}[!t]
\centering
\includegraphics[width=0.98\textwidth]{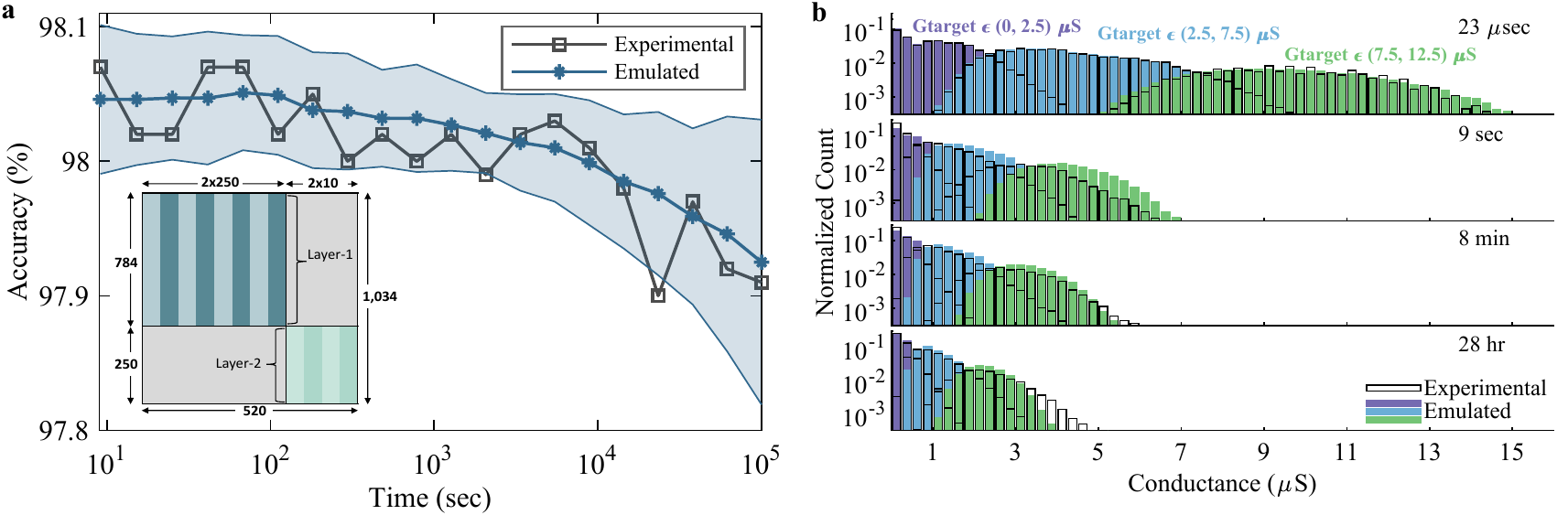}
\vspace{-0.10cm}
\caption{Neural network inference results. (a) Evolution of accuracy over time from PCM experimental results compared with the mean behavior of emulated results and their shaded region representing one standard deviation. The inset presents the mapping scheme of the network's layers in a single emulated crossbar. (b) Experimental and emulated conductance distribution of the neural network's encoded weights for three different conductance ranges.} \label{fig:PCM_Mnist_Results}
\vspace{-0.55cm}
\end{figure*}

The PCM cell emulation model was utilized for emulating a multi-cell PCM crossbar architecture, as shown in Fig. \ref{fig:PCM_Crossbar}a. The depicted architecture uses NxN PCM emulated cells, where each one has its own conductance and drift exponent model parameter. Also, all cells are supplied with samples from the same $1/f$ noise generator, but each emulated cell is fed with a different instantiation of noise samples.

A sequential execution of N dot-products in the emulator can be used to simulate the matrix-vector multiplication of a NxN crossbar in hardware. For our scenario, each element-wise multiplication of the dot-product is performed by one emulated PCM cell. The element-wise multiplication can be used to implement a k-element dot-product in a column of the crossbar, where the k-factor depends on the available hardware resources. Thus, a dot-product with a dimension greater than the k-factor can be achieved by executing its operation several times with the addition of the partial results using a tree structure of adders and an accumulator \cite{petropoulos2018accurate, petropoulos2018versatile}.

The components of the system, which implements the emulated crossbar design, are shown in Fig. \ref{fig:PCM_Crossbar}b. The emulated PCM crossbar consists of two dedicated DRAM memories for storing the crossbar conductances and drift coefficients, while another DRAM is used for storing pre-generated $1/f$ noise samples. Two dedicated data mover engines (HostDRAM - PCM Crossbar) are used for high speed (8 GBps) transfers of data between the crossbar and the server's memory. A PCM crossbar adaptation layer is used for encoding weights to conductances, transforming vector data to voltage values and also decoding the resulted current values. Also, it contains nonlinearity functional blocks (i.e., RELU, sigmoid, tanh) for neural network applications. In addition, the system incorporates a soft-CPU for initialization and control. The soft-CPU interacts with a host application using a dedicated device driver and descriptor-structured data transfers. The PCM crossbar emulator has been implemented on a Kintex UltraScale FPGA and has been tested on a high-end Xeon server.

For a matrix-vector multiplication scenario, the host initiates requests for downloading, conductance encoding, and conductance data storing to the FPGA's DRAM memories. Also, it initializes a dedicated DRAM area with $1/f$ noise samples. In order to start the emulated crossbar for matrix-vector operations, three main procedures are performed: (a) vector data are received through the host, passing by the PCM crossbar adaptation layer and transformed to voltages, (b) concurrently, the conductance matrix is loaded from the DRAMs to the crossbar along with $1/f$ noise values, and the computation is started, (c) finally, the resulted currents are processed by the adaptation layer and then uploaded to the host's DRAM. Such a versatile process can operate in a pipelined fashion using several crossbars.

\vspace{-0.10cm}
\section{Neural Network Inference using the emulated PCM Crossbar}\label{sec:nn}
\vspace{-0.08cm}
For the evaluation of the emulated PCM crossbar, we considered the task of MNIST handwritten digit recognition. For that purpose, we compared emulated, and experimental inference results from PCM arrays over time for a fully-connected neural network with two layers. The network dimensions are 784-250-10, and it was trained in software using single-precision floating-point weights. Next, the trained weights were iteratively programmed to conductance values on the PCM prototype chip, utilizing approximately 400,000 PCM devices. These weights are linearly mapped to conductance values. A differential PCM configuration is used for each synapse where one device denotes the positive part of the weight, and the other device denotes the negative part of the weight. According to the sign of the weight, one device of each differential pair is set close to $\unit[0]{\mu}$S. We use the programmed conductance values of PCM devices at 23 $\mu$sec as the emulator's initial state \cite{Y2019sebastianVLSI}. The subsequent conductance values of later time steps are determined with model parameters (i.e., drift exponent, $1/f$ variance $Q$ factor). For simplicity, we adopt the parameters used to emulate the behavior of devices with target conductance of $\unit[5]{\mu}$S in Section II ($\bar \nu = 0.06$, $\sigma_{\nu} = 0.02$, $Q = 4\times10^{-4}$). Note that the target conductances representing the network weights are mostly contained within the range of 0 to $\unit[5]{\mu}$S.

In Fig. \ref{fig:PCM_Mnist_Results}a, we present accuracy results of neural network inference for a time period greater than 27 hours. The evolution of the mean accuracy over time from the experiment is well captured by the emulator. Additionally, to further verify our model regarding the conductance drift and noise, we show the evolution of the network's weight distribution encoded to conductances. As depicted in Fig. \ref{fig:PCM_Mnist_Results}b, the emulated results capture well the temporal evolution of the conductance distributions for different target conductance states.

For this inference application, both weight layers of the neural network were emulated in a single crossbar in a pipelined fashion. This is achieved by using a crossbar size of 1034x520, to fit both layer dimensions, with redundant cells (zero conductance) in appropriate places of the weights' matrix (see Fig. \ref{fig:PCM_Mnist_Results}a inset). With this mapping approach, our emulator achieves a processing rate of 8.8 kilo-images per second and $\unit[227]{\mu}$sec latency.

\vspace{-0.15cm}
\section{Conclusion}
\vspace{-0.08cm}
In this work, we presented an accurate FPGA-based hardware emulator for phase-change memory that captures the key physical attributes such as temporal drift of conductance values as well as $1/f$ noise. The PCM cell emulator and its extension to the PCM crossbar emulator were experimentally validated using a prototype PCM array based on a deep learning inference hardware experiment that involves approximately 400,000 PCM devices. The presented hardware emulator can be a powerful tool for the exploration of in-memory computing and its applications. This approach is scalable to larger networks and more complex problems, while the application domain is not restricted to neural network inference as this emulator can benefit other in-memory computing scenarios.

\bibliographystyle{IEEEtran}
\bibliography{cites}
\balance
\end{document}